\documentclass[
superscriptaddress,
nofootinbib,
twocolumn,
 amsmath,amssymb,
 aps,
pra,
notitlepage,
longbibliography
]{revtex4-1}

\usepackage{dcolumn}% Align table columns on decimal point
\usepackage{bm}% bold math
\usepackage{epsfig}
\usepackage{graphicx}
\usepackage{latexsym}
\usepackage{amsfonts}
\usepackage{setspace}
\usepackage{graphicx}
\usepackage{amsmath}
\usepackage{verbatim}
\usepackage{color}
\usepackage{SIunits}
\usepackage{hyperref}
\usepackage{chemfig}
\usepackage{braket}
\usepackage{float}
\usepackage{multirow}
\usepackage{bigstrut}
\usepackage{color}
\newcommand{\be}{\begin{equation}}
\newcommand{\ee}{\end{equation}}
\newcommand{\ba}{\begin{array}}
\newcommand{\ea}{\end{array}}
\newcommand{\bqa}{\begin{eqnarray}}
\newcommand{\eqa}{\end{eqnarray}}

\begin{document}

\title{Observation of phonon trapping in the continuum with topological charges}

\author{Hao Tong} 
\thanks{These authors contributed equally to this work.}
\affiliation{Holonyak Micro and Nanotechnology Laboratory and Department of Electrical and Computer Engineering, University of Illinois at Urbana-Champaign, Urbana, IL 61801 USA}
\affiliation{Illinois Quantum Information Science and Technology Center, University of Illinois at Urbana-Champaign, Urbana, IL 61801 USA}
\author{Shengyan Liu} 
\thanks{These authors contributed equally to this work.}
\affiliation{Holonyak Micro and Nanotechnology Laboratory and Department of Electrical and Computer Engineering, University of Illinois at Urbana-Champaign, Urbana, IL 61801 USA}
\affiliation{Illinois Quantum Information Science and Technology Center, University of Illinois at Urbana-Champaign, Urbana, IL 61801 USA}
\author{Mengdi Zhao} 
\affiliation{Holonyak Micro and Nanotechnology Laboratory and Department of Electrical and Computer Engineering, University of Illinois at Urbana-Champaign, Urbana, IL 61801 USA}
\affiliation{Illinois Quantum Information Science and Technology Center, University of Illinois at Urbana-Champaign, Urbana, IL 61801 USA}
\author{Kejie Fang} 
\email{kfang3@illinois.edu}
\affiliation{Holonyak Micro and Nanotechnology Laboratory and Department of Electrical and Computer Engineering, University of Illinois at Urbana-Champaign, Urbana, IL 61801 USA}
\affiliation{Illinois Quantum Information Science and Technology Center, University of Illinois at Urbana-Champaign, Urbana, IL 61801 USA}

\begin{abstract} 
Phonon trapping has an immense impact in many areas of science and technology, from the antennas of interferometric gravitational wave detectors to chip-scale quantum micro- and nano-mechanical oscillators. It usually relies on the mechanical suspension--an approach, while isolating selected vibrational modes, leads to serious drawbacks for interrogation of the trapped phonons, including limited heat capacity and excess noises via measurements. To circumvent these constraints, we realize a new paradigm of phonon trapping using mechanical bound states in the continuum (BICs) with topological features and conducted an in-depth characterization of the mechanical losses both at room and cryogenic temperatures. Our findings of mechanical BICs combining the microwave frequency and macroscopic size unveil a unique platform for realizing mechanical oscillators in both classical and quantum regimes. The paradigm of mechanical BICs might lead to unprecedented sensing modalities for applications such as rare-event searches and the exploration of the foundations of quantum mechanics in unreached parameter spaces.

\end{abstract}
%\pacs{}

\maketitle

Phonon trapping in low-dissipative mechanical systems, such as pristine piezoelectric crystals \cite{vig1994quartz} and bulk/surface acoustic wave resonators \cite{lakin2003review}, has played a crucial role in widespread technologies, including timekeeping and microwave signal processing. In principle, trapping phonons, i.e., excitations of crystal lattice vibrations, can be achieved in detached structures, in contrast to photons which permeate even in the vacuum. Nonetheless, complete mechanical isolation from the environment is unfeasible with common practices and trapping phonons with the ultimate lifetime is an outstanding challenge. Recently, phonon-loss mechanisms and means to enhance mechanical coherence have attracted intensive studies driven by the incentive to scale macroscopic mechanical resonators to micro- and nano-scales \cite{cleland2002noise} for realizing, for example, ultrasensitive mass sensors \cite{li2007ultra,jensen2008atomic} and quantum mechanical oscillators \cite{o2010quantum,teufel2011sideband,chan2011laser}. However, while trapping phonons in isolated vibrational modes, suspended structures with reduced lateral dimensions cause poor thermalization and excess noises when mechanical oscillators are coupled to external probes. These issues are more pressing when coupling optical photons with the trapped phonons, precluding continuous measurements of quantum mechanical oscillators at cryogenic temperatures \cite{meenehan2014silicon,higginbotham2018harnessing} and preparation of high-fidelity quantum mechanical states in ambient conditions \cite{purdy2017quantum,sudhir2017quantum}. 

\begin{figure}[!htb]
\centering
\includegraphics[width =\linewidth]{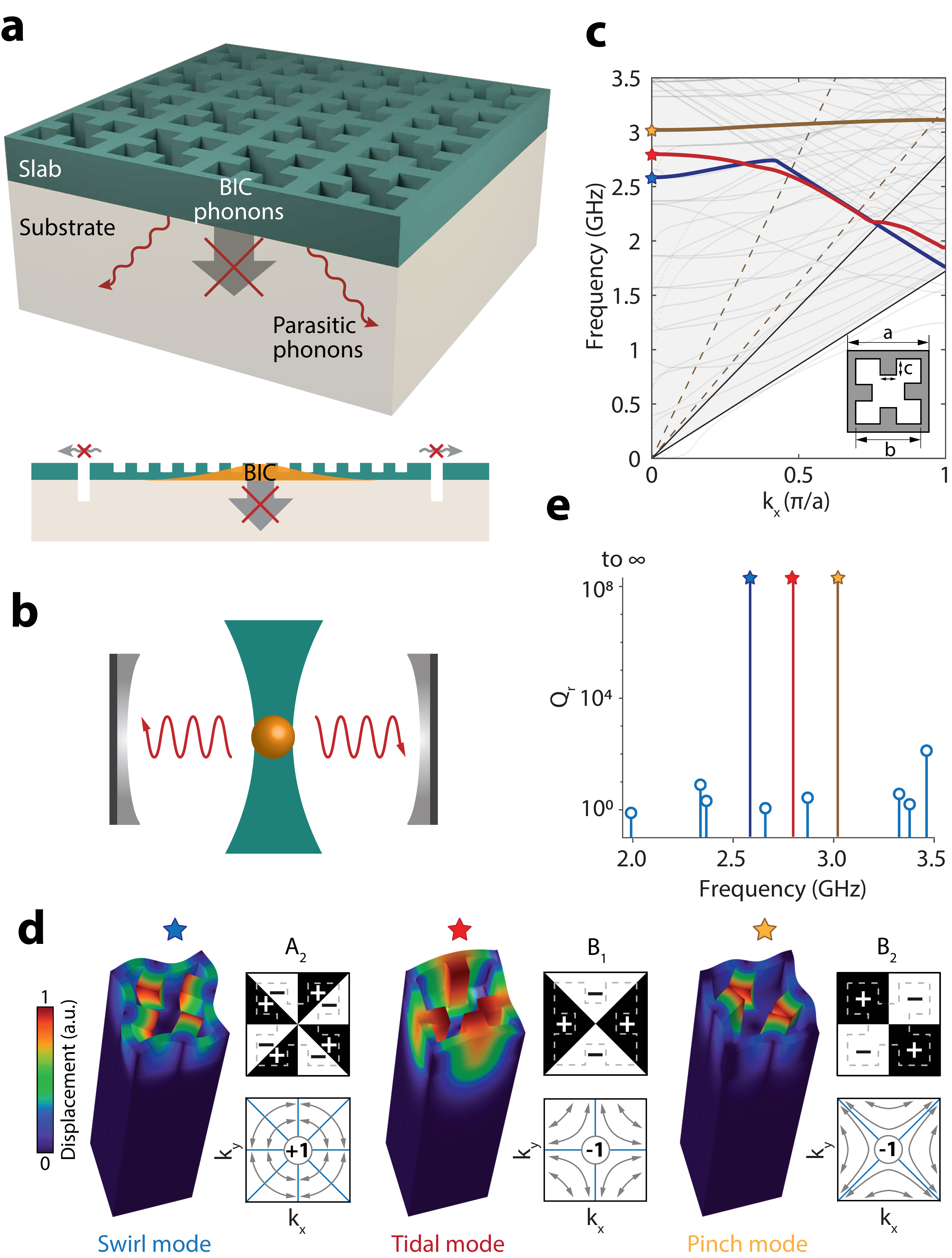}
\caption{\textbf{Phonon trapping via mechanical BICs}. \textbf{a}, Schematic diagram of a slab-on-substrate PnC which traps selected phonons via mechanical BICs while facilitates dissipation of parasitic phonons via the substrate. \textbf{b}, The mechanical BIC architecture resembles levitated particles. \textbf{c}, Calculated mechanical bandstructure of an AlN-on-oxide PnC with parameters given in the text. The solid(dashed) lines correspond to the sound lines of transverse and longitudinal waves of SiO$_2$(AlN), respectively. Three mechanical BICs at the $\Gamma$ point and the associated mechanical bands are highlighted. \textbf{d}, Simulated total displacement of the three BIC modes (swirl, tidal, and pinch) and their group representation under $C_{4v}$ symmetry. Also shown is the far-field transverse polarization of Bloch modes of the same band in the vicinity of $\Gamma$ point, identifying the topological charge of mechanical BICs. Blue lines are the nodal lines of far-field longitudinal polarization. a.u., arbitrary units. \textbf{e}, Simulated radiative quality factor of mechanical modes at the $\Gamma$ point. 
}
\label{fig:1}
\end{figure}

Here we show a novel approach for phonon trapping in a chip-scale architecture via bound states in the continuum (BICs), which is distinctive to the method of mechanical bandgap engineering widely used in suspended structures \cite{eichenfield2009optomechanical,tsaturyan2017ultracoherent,ghadimi2018elastic}. Originally conceived in peculiar quantum mechanical potentials \cite{vonNeumann1929}, BICs are non-radiative states yet spectrally overlapping with the continuum, because of symmetry incompatibility with the radiative modes \cite{ochiai2001dispersion} or accidental radiation amplitude cancellation \cite{yang2014analytical}. Recently, both types of optical BICs have been observed in two-dimensional photonic crystal slabs \cite{lee2012observation, hsu2013observation, kodigala2017lasing}, rendering macroscopic thin-film optical resonators with a quality factor comparable to those of microcavities \cite{jin2019topologically}. Inspired by this, we propose mechanical BICs realized in slab-on-substrate phononic crystals (PnCs), as illustrated in Fig. \ref{fig:1}a, where phonons in the BIC mode are trapped in the unreleased slab without coupling into the substrate, irrespective of the acoustic impedance of the slab and substrate materials, while parasitic phonons are dissipated via the substrate. By further introducing voids on the boundary of the PnC, the radiation loss of BIC-phonons thus can be eliminated, resembling optically trapped nanoparticles (Fig. \ref{fig:1}b), but without the need to levitate the structure, leading to unmatched heat capacity. As our experiments and simulations show, both symmetry-induced and accidental mechanical BICs with topological features can be realized in slab-on-substrate PnCs. In contrast to confined surface or slab acoustic waves with finite momentum below the sound line, mechanical BICs with zero wavevectors are able to couple with single optical resonances in two-dimensional periodic structures \cite{lee2012observation, hsu2013observation}, for non-invasive, optical interrogation of the trapped phonons, similar to the prevailing cavity-optomechanical approaches \cite{aspelmeyer2014cavity}.

\begin{figure*}[htb]
\centering
\includegraphics[width =1\linewidth]{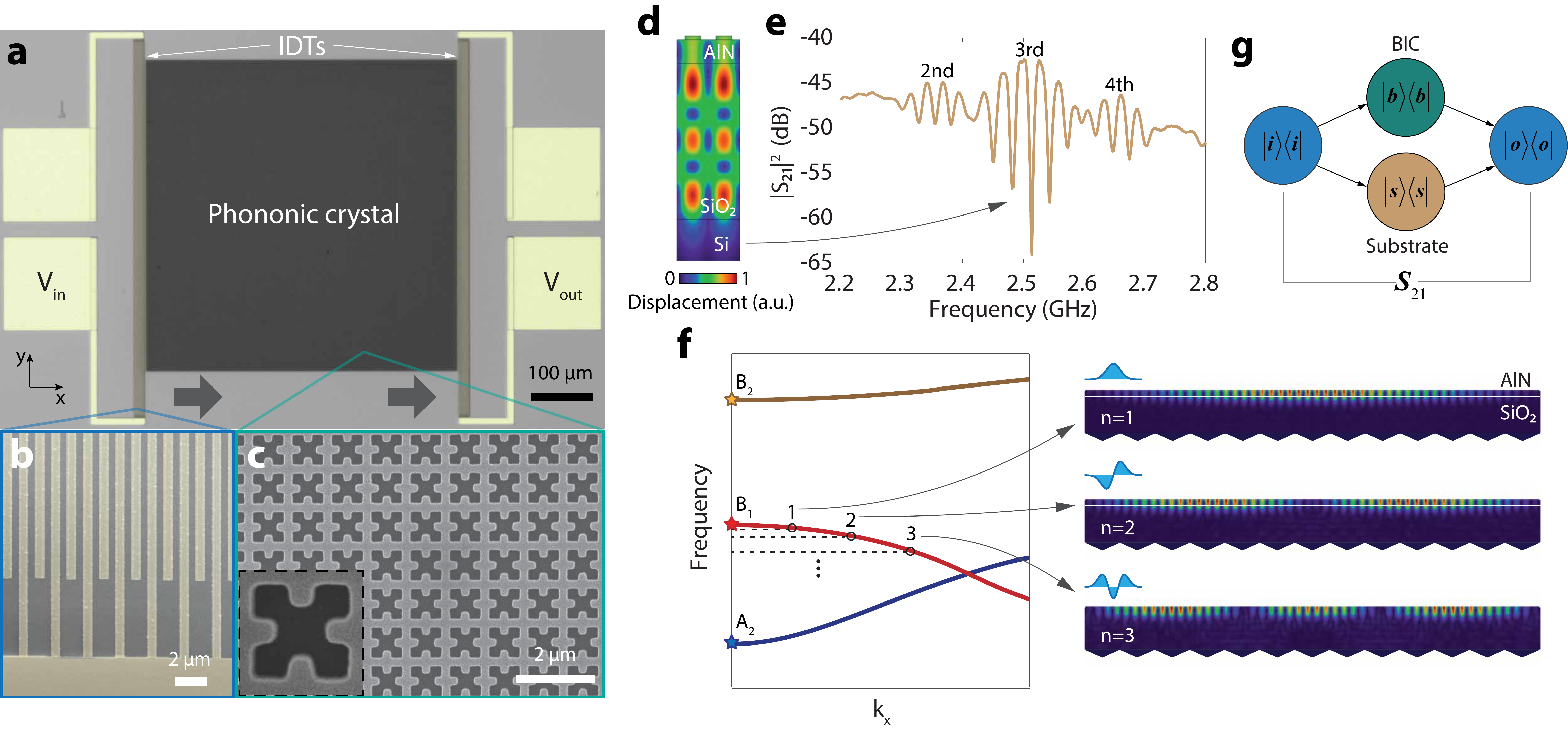}
\caption{\textbf{Device architecture and measurement protocol}. \textbf{a}, False-color optical microscopy image of a device consisting of a PnC with $500\times500$ unit cells and a pair of IDTs with electrodes. \textbf{b} and \textbf{c}, Scanning electron microscopy images of the IDT fingers (\textbf{b}, false-color) and the PnC (\textbf{c}). \textbf{d}, Simulated modal profile of the third-order AlN-SiO$_2$ mode. \textbf{e}, Measured transmission spectrum via a pair of IDTs. \textbf{f}, Formation of standing-wave resonances near the band edge in finite-size PnCs and the simulated modal profile of the first three band-edge modes formed from the $B_1$ BIC in a $y$-periodic PnC. \textbf{g}, Schematic diagram of the measurement protocol.}
\label{fig:2}
\end{figure*}

We designed two-dimensional PnCs which support mechanical BICs in the thin-film aluminum nitride (AlN)-on-oxide structure \cite{zhao2019mechanical}--a material system which otherwise is free of confined microwave-frequency acoustic waves in a uniform slab because of the much larger Young's modulus of AlN than SiO$_2$. Fig. \ref{fig:1}c shows the mechanical bandstructure of a PnC with $C_{4v}$ symmetry and unit-cell dimensions $(a, b, c)=(1000, 800, 200)$ nm in a structure with 600 nm AlN and 4 $\mu$m SiO$_2$ on silicon. 
We numerically found three symmetry-induced mechanical BICs at the $\Gamma$ point, which decouple from both transverse and longitudinal acoustic waves in the substrate, with a frequency and group representation of $(2.58~{\rm GHz}, A_2)$, $(2.80~{\rm GHz}, B_1)$, and $(3.02~{\rm GHz}, B_2)$, respectively, and their modal profile and radiative quality factor shown in Fig. \ref{fig:1}d and e. Despite the fundamentally different nature of acoustic and electromagnetic waves, we found that mechanical BICs are associated with transverse topological charges, i.e., the winding number of the far-field transverse polarization of Bloch modes in the vicinity, similar to the optical counterparts \cite{zhen2014topological}, and yet connected with longitudinal nodal lines in the Brillouin zone where the far-field longitudinal polarization vanishes (Appendix \ref{App:A}), as illustrated in Fig. \ref{fig:1}d. In addition, we also find \emph{robust} existence of accidental mechanical BICs with integer topological charges in slab-on-substrate PnCs despite $z$-symmetry breaking (SI), in contrast to the optical case \cite{yin2019observation}, which can be attributed to the confinement of phonons in solids.

We fabricated the designed PnCs in AlN-on-oxide silicon microchips (see Methods), together with interdigital transducers (IDTs) for piezoelectric actuation of the mechanical BICs (Fig. \ref{fig:2}a-c). Because of the reflection at the oxide-silicon interface, the AlN-SiO$_2$ stack supports weakly confined acoustic modes with energy distributed in the AlN slab \cite{li2015nanophotonic}, which can be piezoelectrically excited by IDTs (Fig. \ref{fig:2}d). We first measured devices without PnCs to characterize the IDT response, with Fig. \ref{fig:2}e showing the measured transmission spectrum via a pair of IDTs separated by 100 $\mu$m, each of which has 20 pairs of fingers and a periodicity of 1.708 $\mu$m. The spectrum exhibits three major envelops whose center frequency corresponds to the second-, third-, and fourth-order AlN-SiO$_2$ modes, while the fringes under the envelope are due to the reflection between the two IDTs forming a weak Fabry-Perot cavity. By tuning the periodicity of IDT fingers, we can adjust the frequency of the AlN-SiO$_2$ modes for them to couple with selected mechanical BICs.

In actual PnCs with $N\times N$ unit cells, the continuous mechanical bands near the band edge discretize into standing-wave modes with constituent momenta $\pm k_{x,y}$ defined relative to the band edge, approximately satisfying the resonance condition $(k_x, k_y)=(n, m)\pi/Na$, $n,m\geq 1$, which we label as the $(n, m)$-th mode. For example, Fig. \ref{fig:2}f shows the simulated modal profile of the first three ($n=1, 2, 3$) band-edge modes near the $\Gamma$ point of the mechanical band associated with the $B_1$ BIC. As illustrated in Fig. \ref{fig:2}g, upon intersecting with the PnC, IDT-excited AlN-SiO$_2$ modes split, where the slab part couples with the BIC standing-wave modes and the rest propagates as a separate substrate mode, resulting in the transmission coefficient $S_{21}$ an averaged mixture of the two parts (see Methods). Since the portion of acoustic energy in the slab is significantly less than that in the substrate for the AlN-SiO$_2$ mode, the transmission spectrum will comprise weak BIC standing-wave resonances on top of IDT fringes. 

\begin{figure}[htb]
\centering
\includegraphics[width =1\linewidth]{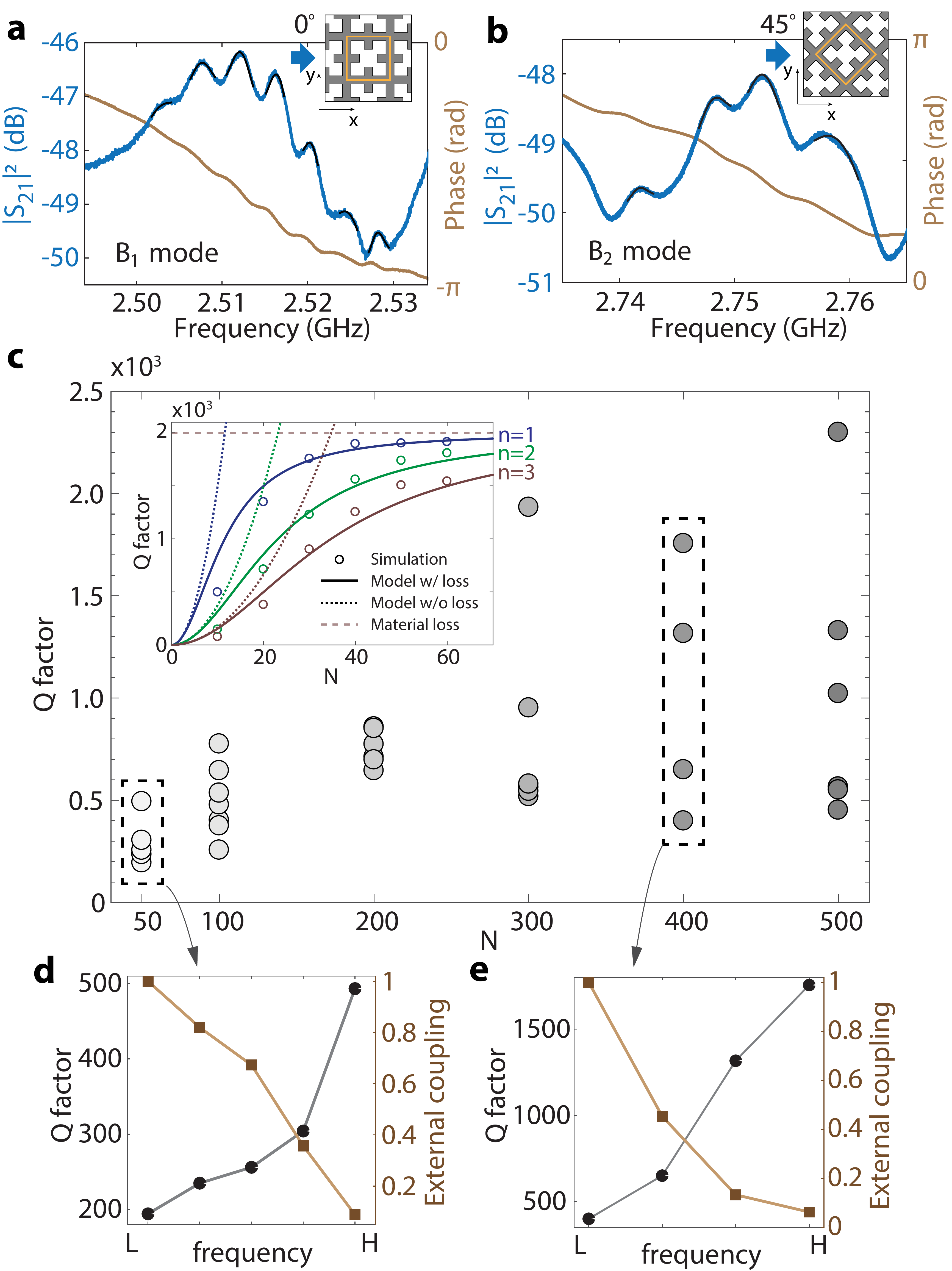}
\caption{\textbf{Room-temperature characterization of mechanical BICs}. \textbf{a} and \textbf{b}, Measured transmission spectrum of a $0^\circ$(\textbf{a})- and $45^\circ$(\textbf{b})-oriented PnC exhibiting $B_1$ and $B_2$ standing-wave modes, respectively. The black curves are the fitted transmission at resonances. \textbf{c}, Observed room-temperature $Q$ factors of $B_1$ standing-wave resonances versus the size of PnCs with $N\times N$ unit cells. The inset shows the $Q$ factor of the first three $B_1$ standing-wave modes in $y$-periodic PnCs by numerical simulation (circles) and theoretical modeling. Material damping of 0.1$\%$ is introduced in the simulation. \textbf{d} and \textbf{e}, $Q$ factor and normalized external coupling ($\omega_m/Q_e$) of the resonances in PnCs for $N=50$ and 400. L, low. H, high.}
\label{fig:3}
\end{figure}

We first made PnCs with the square unit cell aligned with the IDT fingers. Since the IDT-excited acoustic waves are even about the $x$-axis, i.e., the wave propagation direction, only the $B_1$ BIC that is $x$-mirror even will be excited. Fig. \ref{fig:3}a shows the power transmission coefficient of a PnC with $100\times 100$ unit cells measured at room temperature, where several standing-wave modes formed from the $B_1$ BIC are observed in the frequency range of 2.5-2.53 GHz. The frequency discrepancy from the simulation can be attributed to material properties (e.g., the polycrystalline AlN) and fabrication imperfection. To probe the $B_2$ BIC, we rotated the PnC by $45^\circ$ so that the rotated $B_2$ mode is $x$-mirror even and thus can be excited by the incident acoustic wave (Fig. \ref{fig:3}b), while the $B_1$ BIC becomes obscured in this setting. On the other hand, the $A_2$ BIC is inaccessible in either case because of its oddness under both $x$- and diagonal-mirror operations. 

As a new method for phonon trapping, we conducted a comprehensive study of the acoustic loss of mechanical BICs. For the mechanical BIC standing-wave mode, its dissipation can be attributed to three main sources: the radiation loss into the substrate and along the lateral direction, phonon scattering loss caused by the inhomogeneity of the fabricated PnC which breaks spatial periodicity, and material related losses \cite{tabrizian2009effect, kleiman1987two}. In terms of the mechanical quality factor $Q$, i.e., the ratio between the frequency and loss rate, we have
\be\label{Eqn:1}
\frac{1}{Q}\equiv \frac{1}{Q_e}+\frac{1}{Q_i}=\frac{1}{Q_e}+\frac{1}{Q_r}+\frac{1}{Q_s}+\frac{1}{Q_a},
\ee
where $Q_e$ is the external quality factor related to the lateral radiation for probing the BICs, and $Q_i$ is the intrinsic quality factor including components $Q_{r, s, a}$, i.e., the substrate-radiation, phonon-scattering, and material-absorption limited quality factors, respectively. The radiative quality factor $Q_r$ for the $(n, m)$-th BIC standing-wave mode in a $N\times N$ PnC can be modeled as (Appendix \ref{App:B})
\begin{equation}\label{Eqn:2}
{Q_r} \propto \left(\frac{n^2+m^2}{N^2}+\zeta\alpha^{2}\right)^{-1},
\end{equation}
where $\alpha$ is a parameter measuring the symmetry-breaking perturbation of the unit cell and $\zeta>0$ is a constant for a given BIC. Phonon scattering occurs between isofrequency modes, leading to increased radiation of a BIC standing-wave mode according to Fermi's golden rule. For band-edge modes in the vicinity of the $\Gamma$ point, we show that $Q_s$ follows the same scaling rule as $Q_r$ in sufficiently large PnCs (Appendix \ref{App:B}). Finally, we assume that $Q_a$ to be approximately device-size and mode independent at room temperature. 
  
We fabricated a group of $0^\circ$-oriented PnCs with different sizes and observed standing-wave resonances formed from the $B_1$ BIC in each PnC. The fitted total quality factor of these resonances is summarized in Fig. \ref{fig:3}c. For a given size of PnC, the resonances with higher frequency generally have larger $Q$ factor and less external coupling (Fig. \ref{fig:3}d and e), consistent with the fact that these are lower-order modes as the mechanical band associated with the $B_1$ BIC bends downward near the $\Gamma$ point (Fig. \ref{fig:2}f). This is in contrast to the $B_2$ BIC, whose associated band bends upward near the $\Gamma$ point, leading to larger $Q$ factor for the lower frequency resonances (see Fig. \ref{fig:3}b). We also observed an overall trend of increasing of $Q$ factors with larger device sizes, consistent with the scaling rule of Eq. \ref{Eqn:2}. Since in the presence of material damping all standing-wave modes will have similar material-absorption-limited $Q$ factors in sufficiently large PnCs, we conclude that the observed resonances at room temperature are higher-order modes with significant radiation and scattering losses, while the lower-order modes with the more confined modal profile are obscured by the dominant substrate transmission background because of the much weaker external coupling comparing to their intrinsic losses.

\begin{figure}
\centering
\includegraphics[width =\linewidth]{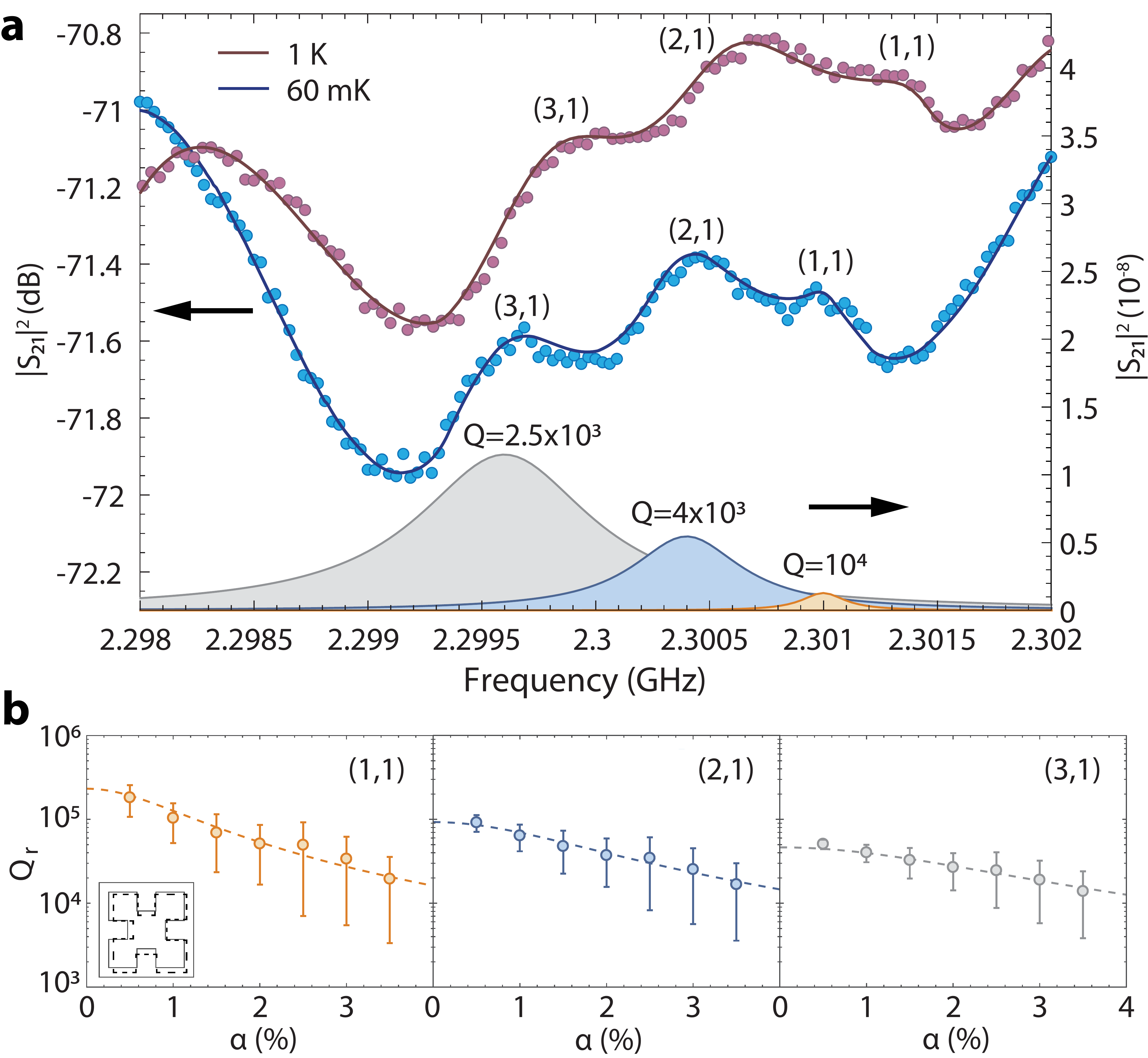}
\caption{\textbf{Radiative quality factor of mechanical BICs at cryogenic temperatures}. \textbf{a}, Transmission spectrum of a PnC with $200\times200$ unit cells measured at 1 K and 60 mK, showing lower-order $B_1$-BIC standing-wave resonances. The right axis shows the transmission via the three resonances at 60 mK by removing the substrate background. \textbf{b}, Simulated radiative quality factor of the three BIC modes using unit cells with disorders in dimensions $b'$s and $c'$s. $\alpha$ is the standard deviation of relative variation. Dashed lines are global fitting using the model of Eq. \ref{Eqn:2}.
}
\label{fig:4}
\end{figure}

To reveal the actual radiation loss of the lower-order BIC standing-wave resonances, we measured the sample at cryogenic temperatures, where material-absorption losses are suppressed. For example, in a PnC with $200\times200$ unit cells, we observed several lower-order $B_1$-BIC standing-wave resonances emerging out of the substrate background (Fig. \ref{fig:4}a). At $T=1$ K, the three resonances, from high to low frequency, have $Q$ factors of $5.2\times10^3$, $2.8\times10^3$, and $2.2\times10^3$, respectively. At $T=60$ mK, the $Q$ factors further increase to $1.01\times10^4$, $4.0\times10^3$, and $2.5\times10^3$, respectively. We found that the mechanical dissipation at 60 mK is dominated by radiation and scattering losses, while the two-level system driven damping \cite{kleiman1987two,maccabe2019phononic}, which is the main anelastic loss at this temperature, is at least three orders weaker (Appendix \ref{App:B}). As a result, the ratio between the $Q$ factor of these resonances at $60$ mK matches well to the scaling rule of Eq. \ref{Eqn:2}, from which we determined the mode order of the three resonances to be $(1,1)$, $(2,1)$, and $(3,1)$, respectively (Appendix \ref{App:B}). We simulated the radiative quality factor of these band-edge modes in the presence of structural disorders as shown in Fig. \ref{fig:4}b, which is comparable to the measured $Q$ factors after taking into account of the estimated scattering loss. Based on these results and the scaling rule of Eq. \ref{Eqn:2}, we expect mechanical BICs with $Q\geq10^7$ can be realized in millimeter-scale PnCs with improved fabrication and material quality, such as using single crystalline silicon \cite{maccabe2019phononic} or epitaxially grown materials \cite{forsch2020microwave,schneider2019optomechanics,ghorbel2019optomechanical}. The mechanical quality factor can also be further improved using the technique of topological charge merging to suppress scattering losses  \cite{jin2019topologically}.

In summary, we have shown a new paradigm of phonon trapping via mechanical BICs in slab-on-substrate PnCs, revealing a fresh ground for studying BIC physics with unique features. The mechanical BIC is expected to enable a new breed of quantum mechanical oscillators with substrate-mediated heat capacity and dissipation of measurement-induced parasitic phonons. We envision several exciting research directions to be enabled by the mechanical BIC architecture. For example, microwave-frequency mechanical BICs in PnCs have an effective mass proportional to $N^2$ that will approach milligrams in centimeter-scale PnCs (note $m_{\textrm{eff}}=0.51$ pg for the 1 $\mu$m-size unit-cell $B_1$-BIC mode), when cryogenically cooled to the quantum ground state ($\bar n_{\textrm{th}}\approx 0.04$ phonons at 5 mK), representing an unparalleled platform for the exploration of macroscopic quantum mechanical effects, including testing wavefunction collapse via continuous spontaneous localization \cite{bassi2013models,forstner2019testing}. The slab-on-substrate architecture also allows integration with heterogeneous structures without introducing extra mechanical dissipations, which might enable unprecedented sensing modalities using high-$Q$ mechanical BICs for direct detection of phonon generation caused by rare physical events in bulk acoustic matrices \cite{knapen2018detection,kurinsky2019diamond,strauss2017gram}.

\noindent\textbf{Methods}:\\
\textbf{Fabrication}: Polycrystalline $c$-axis oriented \chemfig{AlN} thin film is deposited on oxide silicon wafers via a sputtering process with dual cathode S-gun magnetron source.  The AlN PnC is fabricated using electron beam lithography with chemical vapor deposited \chemfig{SiO_2} and ZEP520A as masks, followed by inductively coupled plasma reactive ion etch (ICP-RIE) of oxide using \chemfig{CHF_3} and another ICP-RIE of AlN using \chemfig{BCl_3/Cl_2/Ar}. The chip is dipped in buffered hydrofluoric acid to remove the residual oxide mask of a few nanometers. The IDTs and electrodes are then fabricated using electron beam lithography and lift-off process with 100 nm evaporated  aluminum.\\
\textbf{Measurement}: Room-temperature transmission spectrum was measured using RF probes in contact with the on-chip electrodes. A vector network analyzer (VNA) was used for generating RF signals and receiving transmission spectrum.  For cryogenic temperature device measurements, a printed circuit board is employed as mount and the chip was wire-bonded to it for electrical connection. The mount was placed in the mixing chamber of the dilution refrigerator. Strong attenuation was applied to the RF cables with 5 dBm output power from the VNA. The measurement was conducted in vacuum and low temperature (down to 60 mK).\\
\textbf{Transmission spectrum}: The IDT-excited acoustic wave splits after encountering the PnC and the two parts propagate via the BIC mode in the slab and the SiO$_2$ substrate mode, respectively. The two parts of the acoustic wave experience different losses and local environment and, as a result, develop a relative phase that fluctuates over time.  The transmission coefficient $S_{21}$ can be decomposed as
$S_{21}=S_{s,21}+e^{\delta\varphi}S_{b,21}$, where $S_{s(b),21}$ is the transmission coefficient via the substrate(slab) and $\delta\varphi$ is the relative phase between the two components. The VNA measures the averaged amplitude and phase of $S_{21}$:
\bqa\nonumber
|S_{21}|^2&=&|S_{s,21}|^2+|S_{b,21}|^2+2\langle\cos(\theta+\delta\varphi)\rangle_{\delta\varphi}|S_{s,21}||S_{b,21}|\\\label{Eqn:3}&\approx&|S_{s,21}|^2+|S_{b,21}|^2
\eqa
\be\label{Eqn:4}
{\rm Arg}[S_{21}]\approx {\rm Arg}[S_{s, 21}]+\langle\sin^{-1}\big(\frac{|S_{b,21}|}{|S_{s,21}|}\sin(\theta+\delta\varphi)\big)\rangle_{\delta\varphi},
\ee
where $\theta={\rm Arg}[S_{b, 21}]-{\rm Arg}[S_{s, 21}]$ and $\langle\cdot \rangle_{\delta\varphi}$ means averaging over phase $\delta\varphi$. We have assumed $|\langle\cos(\theta+\delta\varphi)\rangle_{\delta\varphi}|\ll 1$ for quasi-random $\delta\varphi$ in Eq. \ref{Eqn:3} and $|S_{b,21}|\ll |S_{s,21}|$ in Eq. \ref{Eqn:4} based on the simulated acoustic energy distribution of the AlN-SiO$_2$ mode. $S_{b,21}$ can be derived using input-output formalism, 
\be
\frac{db}{dt}=(-i\omega_m-\frac{\gamma}{2})b+\sqrt{\frac{\gamma_e}{4}}b_{\textrm{in}},\quad b_{\textrm{out}}=\sqrt{\frac{\gamma_e}{4}}b,
\ee
which leads to
\be\label{L}
S_{b,21}[\omega]\equiv \frac{b_{\textrm{out}}[\omega]}{b_{\textrm{in}}[\omega]}=\frac{\frac{\gamma_e}{4}}{i(\omega_m-\omega)+\frac{\gamma}{2}},
\ee
where $b$ is the mode amplitude of the BIC, $\gamma\equiv\omega_m/Q$ and $\gamma_e\equiv\omega_m/Q_e$. Fitting of the measured transmission spectrum thus uses Eq. \ref{Eqn:3} with $|S_{b,21}|^2$ comprising multiple Lorentzian functions. From Eq. \ref{L}, we also inferred $\gamma_e$ shown in Fig. \ref{fig:3} d and e.

%\bibliographystyle{naturemag}
%\bibliographystyle{naturemag_NoURL_allnames}
%\bibliography{./reference}

\vspace{2mm}
\noindent\textbf{Acknowledgements}\\ 
We are grateful to Tian Zhong for the use of his dilution refrigerator. This work is supported by US National Science Foundation under Grant No. ECCS-1809707 and No. ECCS-1944728.

\onecolumngrid
\appendix

\section{Transverse topological charges and accidental mechanical BICs}\label{App:A}

\subsection{Definition of transverse topological charges}
We define a topological charge associated with mechanical BICs, based on the far-field transverse polarization.  We decompose the far-field amplitude as $\mathbf{c} = \mathbf{c}_T + \mathbf{c}_L = {c_T}\mathbf{e}_T + {c_L}\mathbf{e}_L$, where $\mathbf{e}_{T,L}$ are unit vectors along the transverse and longitudinal polarization directions (Fig. \ref{fig:sche}a). It is straightforward to show 
\begin{equation}
\mathbf{c}_L = \frac{{\mathbf{c} \cdot \mathbf{k}_T}}{{\mathbf{e}_L \cdot \mathbf{k}_T}}\mathbf{e}_L, \ \mathbf{c}_T = \mathbf{c}- \mathbf{c}_L,
\end{equation}
where $\mathbf{k}_T\perp\mathbf{e}_T$ is the wavevector of the transverse far-field. 

From the transverse radiation field $\mathbf{c}_T(\mathbf{k})$, we define the topological charge of a Bloch mode to be 
\begin{equation}
q=\frac{1}{2 \pi} \oint_{C} d \mathbf{k} \cdot \nabla_{\mathbf{k}} \theta(\mathbf{k}),
\end{equation}
where $C$ is a path enclosing the Bloch mode in Brillouin zone and $\theta$ is the angle between the major axis of the elliptical polarization and the $k_x$ axis. Here, we define the major axis $\mathbf{A_c}$ to be that of the in-plane polarization vector $\mathbf{c}_{T,\parallel}$ \cite{berry2004index}, i.e., the projection of $\mathbf{c}_{T}$ onto the $x$-$y$ plane,
\begin{equation}
\mathbf{A_c}=\frac{1}{|\sqrt{\mathbf{c}_{T,\parallel} \cdot \mathbf{c}_{T,\parallel}}|} \operatorname{Re}\left[\mathbf{c}_{T,\parallel} \sqrt{\mathbf{c}_{T,\parallel}^{*} \cdot \mathbf{c}_{T,\parallel}^{*}}\right].
\end{equation}

Based on this definition, we calculated the in-plane polarization field distribution and topological charge of the three mechanical BICs at the $\Gamma$ point, as shown in Fig. 1d. All of them are associated with integer charges.

\subsection{Accidental BICs on high-symmetric lines}

Besides the BICs at high-symmetric points, there could be accidental BICs at regular $\bm k$ points in the first Brillouin zone. One argument for this is based on radiation amplitude cancellation.  In the optical case, since electromagnetic fields only have transverse polarization, it is possible to tune the real far-field radiation amplitude $(c_x(k_x, k_y),c_y(k_x, k_y))$ to be zero for some Bloch wavevector $(k_x, k_y)$ in $z$-symmetric structures \cite{hsu2013observation}. However, for the acoustic case, because of the existence of both transverse and longitudinal waves and they have different sound speed, in general it is impossible to tune $(c_x(k_x, k_y),c_y(k_x, k_y), c_z(k_x, k_y))$ to be all zero with only two variables $k_x$ and $k_y$.

It turns out, with some additional symmetry constraints, it is possible to realize accidental mechanical BICs. For this purpose, we consider potential accidental BICs on the high-symmetric lines, i.e., $\Gamma\to X$ and $\Gamma\to M$. For Bloch modes on the high-symmetric lines, they belong to the representation $A'$ and $A''$ of $C_{1h}$ group, which are even and odd under the mirror operation about the line, respectively. For the far-field longitudinal polarization to vanish, the mode has to be odd. This also dictates the far-field transverse polarization to be perpendicular to the mirror plane. As a result, for modes on the high-symmetric lines, their far-field radiation amplitude only has one transverse component, which can be chosen to be real following the argument of Eqs. \ref{SEqn:4} and \ref{SEqn:5}. Since phonons only propagate in solids, thus there is only downward acoustic radiation in the slab-on-substrate structures. Thus, by varying the Bloch wavevector along the high-symmetric line, it is possible to eliminate this only component to realize accidental mechanical BICs. 

\begin{table}[htbp]
	\centering
	\caption{Representation reduction of $C_{4v}$ group along $\Gamma\to X$ and $\Gamma\to M$}
	\begin{tabular}{c|c|c}
		\hline
		\hline
		$C_{4v}$   & $C_{1h}(\Gamma \to X)$ & $C_{1h}(\Gamma \to M)$ \bigstrut\\
		\hline
		$A_1$    & $A'$    & $A'$ \bigstrut\\
		\hline
		$A_2$    & $A''$   & $A''$ \bigstrut\\
		\hline
		$B_1$    & $A'$    & $A''$ \bigstrut\\
		\hline
		$B_2$    & $A''$   & $A'$ \bigstrut\\
		\hline
		$E$     & $A'+A''$ & $A'+A''$ \bigstrut\\
		\hline
	\end{tabular}%
	\label{tab:reduction}
\end{table}%

\begin{figure}[htb]
	\centering
	\includegraphics[width=0.6\linewidth]{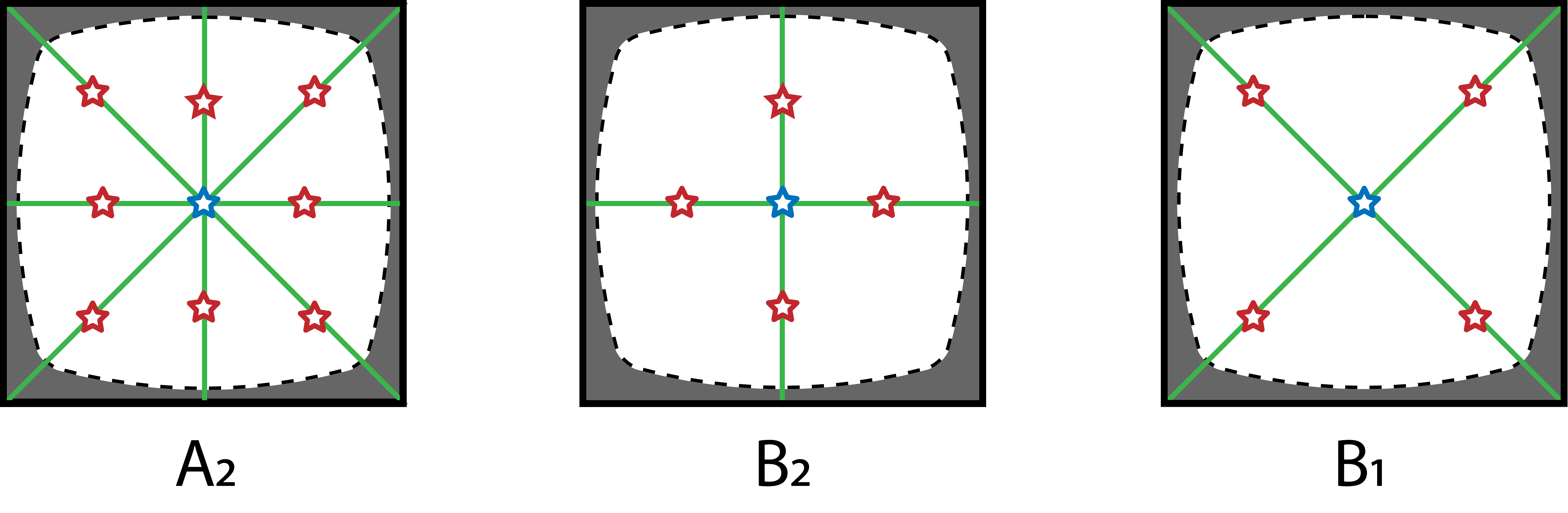}
	\caption{ Possible accidental mechanical BICs (red star) associated with symmetry-induced BICs (blue star) at the $\Gamma$ point.  Green lines indicate the nodal lines of longitudinal radiation amplitude. }
	\label{SFig:bics}
\end{figure}

Based on this argument, we examine possible accidental mechanical BICs in structures with $C_{4v}$ symmetry and their relation to the modes at the $\Gamma$ point. As the wavevector moves away from the $\Gamma$ point along the high-symmetric lines $\Gamma\to X$ and $\Gamma\to M$, the representations of $C_{4v}$ group at the $\Gamma$ point reduces to representations of $C_{1h}$ group, as summarized in Table \ref{tab:reduction}. Only modes with the $A''$ representation, i.e., odd under mirror operation, could possibly be a mechanical BIC. In Fig. \ref{SFig:bics}, we illustrate, specifically for $A_2$, $B_2$, and $B_1$ modes at the $\Gamma$ point which are symmetry-induced BICs, high-symmetric lines along which they could evolve to accidental BICs.

\subsection{Numerical demonstration of accidental BICs}
\begin{figure}
	\centering
	\includegraphics[width=0.9\linewidth]{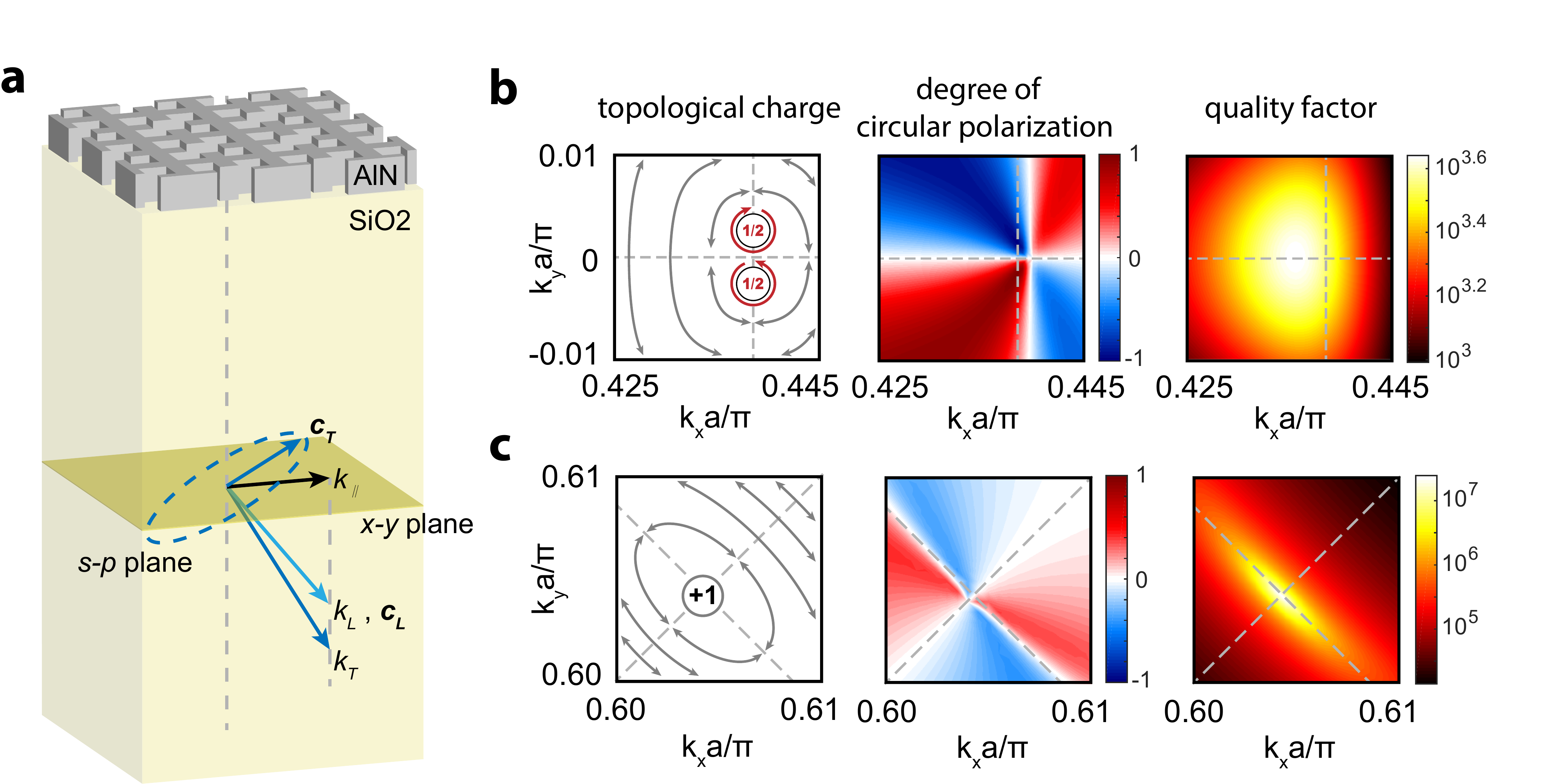}
	\caption{\textbf{a}, Schematics of decomposition of far-field radiation amplitude. \textbf{b}, Half-charge high-$Q$ modes along the $\Gamma\to X$ line. Left panel: transverse polarization vector field and topological charge. Middle panel: degree of circular polarization. Right panel: quality factor. \textbf{c}, Accidental BIC with integer topological charge on the $\Gamma\to M$ line. }
	\label{fig:sche}
\end{figure}

For the structure studied in the main text, we swept the mechanical bands along the high-symmetric lines, $\Gamma\to X$ and $\Gamma\to M$, and find high-$Q$ mechanical modes. We also calculated the transverse far-field polarization which reveals the topological properties of these high-$Q$ modes. It turns out that the high-$Q$ mode on the $\Gamma\to X$ line is associated with two half-charges, i.e., $q=1/2$, aside of the $x$-axis (Fig. \ref{fig:sche}b), while the one on the $\Gamma\to M$ line is an accidental BIC with an integer charge (Fig. \ref{fig:sche}c) which is evolved from a mode belong to the $E$ representation at the $\Gamma$ point. 

We also characterized the degree of circular polarization near the high-$Q$ modes. The far-field transverse polarization of the Bloch mode can be decomposed in the circular polarization basis, i.e., 
\be
\mathbf{c}_{T,||}=\alpha_R\mathbf{c}_R+\alpha_L\mathbf{c}_L,
\ee
and the degree of circular polarization is defined as
\begin{equation}
{m_c} =\frac{|\alpha_R|^2-|\alpha_L|^2}{|\alpha_R|^2+|\alpha_L|^2}\in [ - 1,1].
\end{equation}
As seen from Fig. \ref{fig:sche}c, the accidental BIC is located at the intersection of two perpendicular lines with $m_c=0$ corresponding to linear polarizations. The quality factor of the accidental mechanical BIC with non-zero wavevectors is less than the BICs at the $\Gamma$ point. We find this might just be numerical manifestation due to the simulation software COMSOL, whose default acoustic PML/low-reflection boundary conditions do not work perfectly for modes with large transverse momentum, because it turns out that mechanical modes below the acoustic line are also missing ultrahigh $Q$ factors.

We find the accidental BIC on the $\Gamma\to M$ line persists as we vary the thickness of the AlN slab. The robust existence of the accidental BIC in certain structural parameter range can be argued as follows. For a given unit cell structure, we vary one specific parameter, e.g., the thickness of slab $h$. The transverse polarization field $c_T$ perpendicular to the high-symmetric line, which is chosen to be real, is a function of $h$ and $k$ along the high-symmetric line. \emph{If} there exists an accidental BIC for some $h_0$ and $k_0$, i.e., $c_T(h_0, k_0)=0$, and since the zeros of the 2D function $c_T(h, k)$ in general consists of a 1D curve in the $h$-$k$ plane,  one can always find $k'$ satisfying $c_T(h', k')=0$ for arbitrary $h'$ around $h_0$.

%%%%%%%%%%%%%%%%%%%%%%%%%%%%%%%%%%%%%%%%%%

\section{Quality factor of mechanical BICs}\label{App:B}

\subsection{Radiative quality factor: symmetry-preserving structure}
We derive the radiative quality factor of Bloch modes in the vicinity of mechanical BICs at the $\Gamma$ point. The mechanical Bloch modes satisfy the following eigenequation
\begin{equation}\label{SEqn:1}
-\omega^2\rho ({\bf{r}})Q_i({\bf{r}}) = {\partial _j}\left({C_{ijkl}}({\bf{r}}){\partial _k}{Q_l}({\bf{r}})\right),
\end{equation}
where $\rho(\mathbf{r})$ is the density, $C_{ijkl}(\mathbf{r})$ is the elasticity tensor, and $Q_l(\mathbf{r})= {e^{i\mathbf{k}\cdot\mathbf{r}}}u_{\mathbf{k}l}(\mathbf{r})$ is the displacement field of the Bloch mode. For a transversely isotropic material system, such as the AlN-on-oxide system studied in this paper, the elastic tensor takes the form
\begin{equation}
C=\left[\begin{array}{cccccc}C_{1111} & C_{1122} & C_{1133} & 0 & 0 & 0 \\ & C_{1111} & C_{1133} & 0 & 0 & 0 \\ & & C_{3333} & 0 & 0 & 0 \\ & & & C_{2323} & 0 & 0 \\ & & s y m m & & C_{2323} & 0 \\ & & & & & \frac{1}{2}\left(C_{1111}-C_{1122}\right)\end{array}\right].
\end{equation}
When the system possesses $C_{2}^{z} T$ symmetry, i.e., $C_{ijkl}^*(x,y,z) = {C_{ijkl}}( - x, - y,z)$ and ${\rho ^*}(x,y,z) = \rho ( - x, - y,z)$, we find that for an eigenfunction $\mathbf{u}_\mathbf{k}(\mathbf{r})$ of Eq. \ref{SEqn:1}, 
\be
C_{2}^{z} \mathbf{u}_\mathbf{k}^{*}\left(C_{2}^{z} \mathbf{r}\right) = \big(-u_{\mathbf{k},1}^*( - x, - y,z), - u_{\mathbf{k},2}^*( - x, - y,z),u_{\mathbf{k},3}^*( - x, - y,z)\big)
\ee
is also an eigenfunction with the same eigenvalue. Thus, $\mathbf{u}_\mathbf{k}(\mathbf{r})$ and $C_{2}^{z} \mathbf{u}_\mathbf{k}^{*}\left(C_{2}^{z} \mathbf{r}\right) $ must differ at most by an arbitrary phase factor, which can be chosen to be $-1$ for all $\mathbf{k}$, i.e.,
\be\label{SEqn:4}
\big(u_{\mathbf{k},1}(x, y, z), u_{\mathbf{k},2}(x, y, z),u_{\mathbf{k},3}(x, y, z)\big)=\big(u_{\mathbf{k},1}^*( - x, - y,z), u_{\mathbf{k},2}^*( - x, - y,z), -u_{\mathbf{k},3}^*( - x, - y,z)\big).
\ee
The far-field radiation amplitude of the Bloch mode is characterized by the averaged value of $\mathbf{u}_\mathbf{k}(\mathbf{r})$ in the $x-y$ plane, i.e., 
\be\label{SEqn:5}
\left( c_x(\mathbf{k}), c_y(\mathbf{k}),c_z(\mathbf{k}) \right)\equiv\left(\left<u_{\mathbf{k},1}\right>_{xy}, \left<u_{\mathbf{k},2}\right>_{xy}, \left<u_{\mathbf{k},3}\right>_{xy} \right),
\ee
Thus, according to Eq. \ref{SEqn:4}, $c_x(\mathbf{k})$ and $c_y(\mathbf{k})$ are real while $c_z(\mathbf{k})$ is pure-imaginary.

Next we relate the eigenfunction at $\mathbf{k}$ and the eigenfunction at $\mathcal{R}\mathbf{k}$, where $\mathcal{R}$ is an element of the point group of the structure. Since the structure is invariant under transformation $\mathcal{R}$, we know $\hat{O}_{\mathcal{R}} \mathbf{u}_{\mathbf{k}}$ is also an eigenfunction of Eq. \ref{SEqn:1} with Bloch wavevector $\mathcal{R} \mathbf{k}$. In the case without degeneracy, we have \(\hat{O}_{\mathcal{R}} \mathbf{u}_{\mathbf{k}}=\alpha_{\mathbf{k}} \mathbf{u}_{\mathcal{R} \mathbf{k}}\). Since $\mathbf{u}_{\mathbf{k}}$ is chosen to fulfill the realness of $c_{x,y}$ and $ic_z$, and $\hat{O}_{\mathcal{R}}$ does not mix $x,y$-components with $z$-component for a 2-D point group, $\alpha_{\mathbf{k}}$ has to be real. When $\mathbf{u}_{\mathbf{k}}$ is normalized, we obtain $\alpha_{\mathbf{k}}=\pm 1$. Because $\alpha_{\mathbf{k}}$ is also continuous regarding $\mathbf{k}$, it has to be a constant (1 or -1), which thus equals $\alpha_{\Gamma}$. The latter is related to the spatial symmetry of the mode at the $\Gamma$ point, i.e., \(\hat{O}_{\mathcal{R}} \mathbf{u}_{\Gamma}=\alpha_{\Gamma} \mathbf{u}_{\Gamma}\) (since $\mathcal{R}\Gamma=\Gamma$). In conclusion, we have 
\be\label{SEqn:60}
\mathbf{u}_{\mathcal{R}\mathbf{k}}=\alpha_{\Gamma} \hat{O}_{\mathcal{R}} \mathbf{u}_{\mathbf{k}}.
\ee
And the averaged far-field amplitude of Eq. \ref{SEqn:5} obeys the same relation. This result is identical to the optical case \cite{zhen2014topological}.

Using Eq. \ref{SEqn:60}, we can find the asymptotic expression of the far-field amplitude of the Bloch modes in the vicinity of BICs at the $\Gamma$ point. The averaged far-field amplitude of the Bloch modes in the vicinity of BICs can be expanded around $\mathbf{k} = 0$ as:
\begin{equation}\label{SEqn:6}
c_{x}(\mathbf{k})=\sum_{\substack{m,n\ge 0\\m+n\geq 1}} p_{m n} k_{x}^{m} k_{y}^{n}, \  c_{y}(\mathbf{k})=\sum_{\substack{m,n\ge 0\\m+n\geq 1}} q_{m n} k_{x}^{m} k_{y}^{n}, \ c_{z}(\mathbf{k})=\sum_{\substack{m,n\ge 0\\m+n\geq 1}} r_{m n} k_{x}^{m} k_{y}^{n},
\end{equation}
because at $\mathbf{k}=0$, $c_{x}(0)=c_{y}(0)=c_{z}(0)=0$ for BICs. Taking the $B_1$ BIC mode as an example, it is even under mirror operation $\sigma_{x,y}$, i.e.,  $\mathbf{u}_{\sigma_{x,y}\mathbf{k}}=\sigma_{x,y} \mathbf{u}_{\mathbf{k}}$, so we have
\begin{equation}\label{SEqn:8}
\begin{aligned}
&c_{x}\left(k_{x},k_{y}\right)=c_{x}\left(k_{x},-k_{y}\right)=-c_{x}\left(-k_{x},k_{y}\right),\\
& c_{y}\left(k_{x},k_{y}\right)=-c_{y}\left(k_{x},-k_{y}\right)=c_{y}\left(-k_{x},k_{y}\right),\\
& c_{z}\left(k_{x},k_{y}\right)=c_{z}\left(k_{x},-k_{y}\right)=c_{z}\left(-k_{x},k_{y}\right).
\end{aligned}
\end{equation}
As a result, only those terms in Eq. \ref{SEqn:6} respecting the symmetry constraint of Eq. \ref{SEqn:8} will be present, i.e., to the leading order
\begin{equation}\label{SEqn:9}
\begin{aligned}
&c_{x}\left(k_{x},k_{y}\right)=p_{1} k_{x}+p_{2} k_{x}^{3}+p_{3} k_{x} k_{y}^{2}+O\left(k^{5}\right),\\
&c_{y}\left(k_{x},k_{y}\right)=q_{1} k_{y}+q_{2} k_{y}^{3}+q_{3} k_{y} k_{x}^{2}+O\left(k^{5}\right),\\
&c_{z}\left(k_{x},k_{y}\right)=r_{1} k_{x}^2+r_{2} k_{y}^2+r_{3} k_{x}^4+r_{4} k_{y}^4+r_{5} k_{x}^2 k_{y}^2+O\left(k^{6}\right).
\end{aligned}
\end{equation}
The $B_1$ BIC is also odd under $\frac{\pi}{2}$ rotation around the $z$-axis, i.e., $\mathbf{u}_{\mathcal{R}_{\frac{\pi}{2}} \mathbf{k}}=-\mathcal{R}_{\frac{\pi}{2}}\mathbf{u}_\mathbf{k}$, which leads to 
\begin{equation}\label{SEqn:11}
\begin{aligned}
& {c_x}\left( { - {k_y},{k_x}} \right)={c_y}\left( {{k_x},{k_y}} \right), \\
& {c_y}\left( { - {k_y},{k_x}} \right)=- {c_x}\left( {{k_x},{k_y}} \right), \\
& {c_z}\left( { - {k_y},{k_x}} \right)=-{c_z}\left( {{k_x},{k_y}} \right).
\end{aligned}
\end{equation}
Eq. \ref{SEqn:11} further constrains the coefficients in Eq. \ref{SEqn:9}, leading to $p_{1,2,3} =- −q_{1,2,3}$, $r_{1,3}=-r_{2,4}$, and $r_5=0$. In conclusion, the radiation field of the Bloch modes near the $B_1$ mechanical BIC is given by
\begin{equation}\label{SEqn:12}
\begin{aligned}
{c_x}\left(k_{x},k_{y}\right) &= {p_1}{k_x} + {p_2}k_x^3 + {p_3}{k_x}k_y^2 + O\left( {{k^5}} \right),\\
{c_y}\left(k_{x},k_{y}\right) &=  - {p_1}{k_y} - {p_2}k_y^3 - {p_3}{k_y}k_x^2 + O\left( {{k^5}} \right),\\
{c_z}\left(k_{x},k_{y}\right) &= {r_1}(k_x^2 - k_y^2) + {r_3}(k_x^4 - k_y^4) + O\left( {{k^6}} \right).
\end{aligned}
\end{equation}
Similarly, we find for Bloch modes in the vicinity of the $B_2$ BIC,
\begin{equation}\label{SEqn:13}
\begin{aligned}
c_{x}\left(k_{x}, k_{y}\right)&=p_{1} k_{y}+p_{2} k_{y}^{3}+p_{3} k_{x}^{2} k_{y}+O\left(k^{5}\right), \\ c_{y}\left(k_{x}, k_{y}\right)&=p_{1} k_{x}+p_{2} k_{x}^{3}+p_{3} k_{x} k_{y}^{2}+O\left(k^{5}\right), \\ c_{z}\left(k_{x}, k_{y}\right)&=r_{1} k_{x} k_{y}+r_{2} k_{x} k_{y}\left(k_{x}^{2}+k_{y}^{2}\right)+O\left(k^{6}\right),
\end{aligned}
\end{equation}
and of the $A_2$ BIC,
\begin{equation}\label{SEqn:14}
\begin{aligned}
c_{x}\left(k_{x}, k_{y}\right)&=p_{1} k_{y}+p_{2} k_{y}^{3}+p_{3} k_{x}^{2} k_{y}+O\left(k^{5}\right), \\ c_{y}\left(k_{x}, k_{y}\right)&=-p_{1} k_{x}-p_{2} k_{x}^{3}-p_{3} k_{x} k_{y}^{2}+O\left(k^{5}\right), \\ c_{z}\left(k_{x}, k_{y}\right)&=r_{2} k_{x} k_{y}\left(k_{y}^{2}-k_{x}^{2}\right)+O\left(k^{6}\right).
\end{aligned}
\end{equation}

\begin{figure}[htb]
	\centering
	\includegraphics[width=0.7\linewidth]{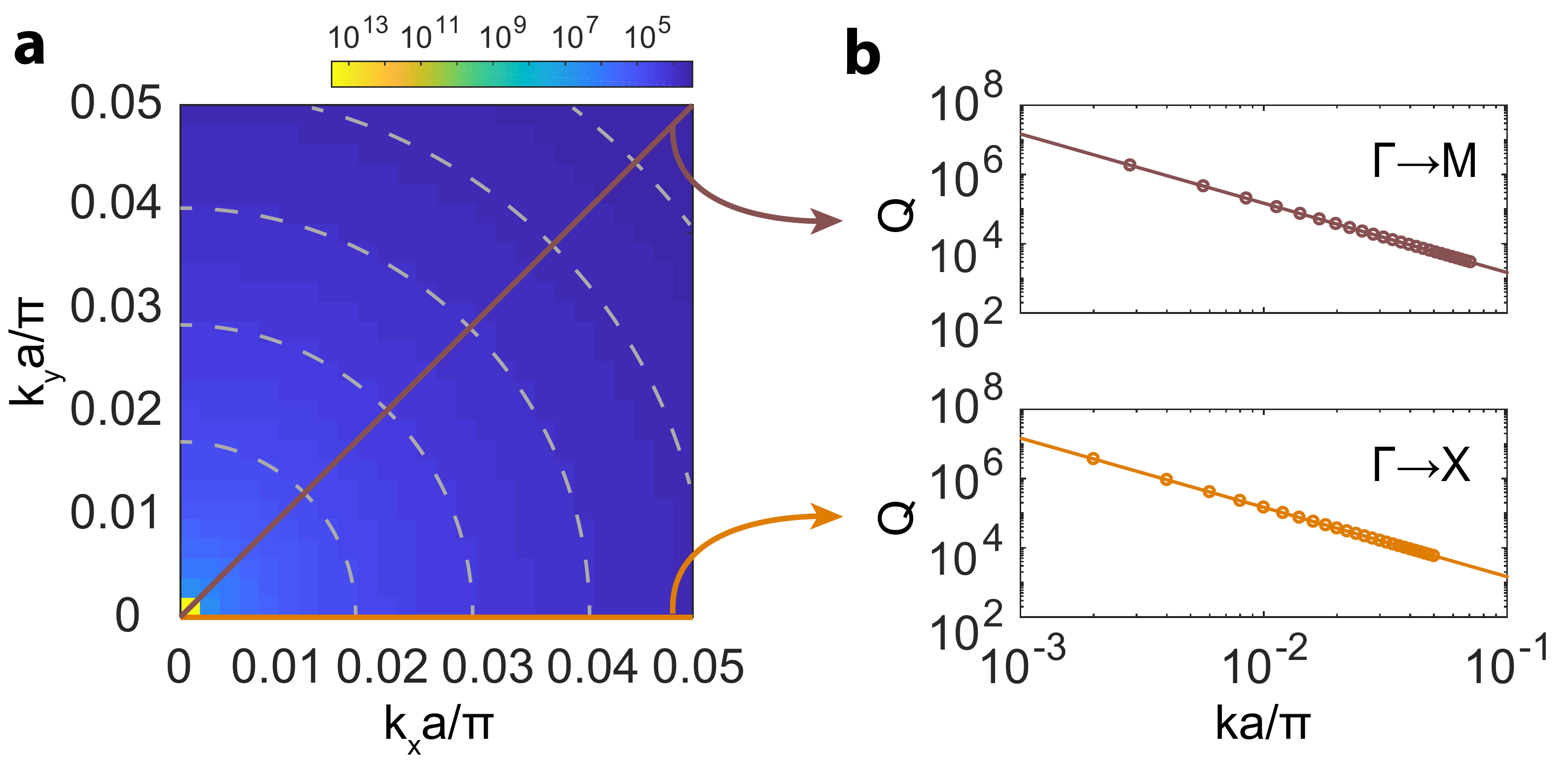}
	\caption{ \textbf{a}, Simulated radiative quality factor of Bloch modes in the vicinity of $B_1$ BIC at the $\Gamma$ point. Dashed lines are equi-$Q$ factor contours. \textbf{b}, Two cut lines along $\Gamma-M$ and $\Gamma-X$ direction. Solid lines show the scaling rule of $Q_r \propto 1/k^2$.}
	\label{SFig:1}
\end{figure}

The radiative quality factor of the Bloch mode is inversely proportional to the outflow acoustic energy flux. The acoustic energy flux $\mathbf{P}$ is defined as  \cite{synge1956flux}
\begin{equation}
\mathbf{P} =  - \bm{\tau} \dot{\mathbf{Q}}^*,
\end{equation}
where the stress tensor $\tau_{ij}$ is given by Hooke's law: ${\tau _{ij}} = {c_{ijkl}}\left( {{\partial _k}{Q_l} + {\partial _l}{Q_k}} \right)/2$. Thus, the $z$-component of the energy flux in a transversely isotropic medium is given by
\begin{equation}
{P_z} =  - \lambda \dot Q^*_x{\partial _z}Q_x - \lambda \dot Q^*_y{\partial _z}Q_y - (\lambda  + 2\mu )\dot Q^*_z{\partial _z}Q_z,
\end{equation}
where $\lambda$ and $\mu$ is Lam\'{e}'s first and second parameter, respectively. For Bloch modes near the $\Gamma$ point,  
\begin{equation}
{P_z} \approx   \lambda\frac{\omega^2}{v_T}  (|c_x|^2+|c_y|^2) +(\lambda  + 2\mu )\frac{\omega^2}{v_L}  |c_z|^2,
\end{equation}
where $\omega$ is the frequency of the Bloch mode and ${v_{T(L)}}$ is the sound speed of the transverse(longitudinal) plane-wave. Since $\omega=\omega_0+O(k)$, the dependence of radiative quality factor $Q_r$ on $\mathbf{k}$, to the leading order, is found to be
\begin{equation}\label{SEqn:Qc}
Q_r \propto \frac{1}{P_z}   \propto \frac{1}{ |c_x|^2+|c_y|^2 +\beta |c_z|^2},
\end{equation}
where $\beta = (\lambda  + 2\mu ){v_T}/\lambda {v_L}$, which in conjunction with Eqs. \ref{SEqn:12}, \ref{SEqn:13}, or \ref{SEqn:14} yields
\begin{equation}
Q_r \propto \frac{1}{k_{x}^{2}+k_{y}^{2}},
\end{equation}
where we have assumed the leading terms in $c_z$ is much smaller than the leading terms in $c_{x,y}$. This is confirmed with the numerical simulation shown in Fig. \ref{SFig:1}.

\subsection{Radiative quality factor: symmetry-breaking perturbations}

When the unit cell acquires symmetry-breaking perturbations, the mechanical BICs at the $\Gamma$ point couple into the radiation continuum and become quasi-BICs with finite radiative quality factor. We assume the perturbation can be characterized by a perturbation parameter $\alpha$. The far-field amplitude of the quasi-BIC now contains all the terms of Eq. \ref{SEqn:6} in addition to constants, with the coefficients of these terms being functions of $\alpha$. However, to the leading order of both $k$ and $\alpha$, the far-field amplitude can be approximated by only a few terms. For example, for the $B_1$ quasi-BIC, we have
\begin{equation}
\begin{aligned}
{c_x}\left(k_{x},k_{y}\right) &\approx p_0\alpha+{p_1}{k_x},\\
{c_y}\left(k_{x},k_{y}\right) &\approx q_0\alpha- {p_1}{k_y},\\
{c_z}\left(k_{x},k_{y}\right) &\approx ir_0\alpha,
\end{aligned}
\end{equation}
which, according to Eq. \ref{SEqn:Qc}, leads to 
\bqa
Q_r &\propto& \frac{1}{(p_0\alpha+p_1k_x)^2+(q_0\alpha-p_1k_y)^2+\beta r_0^2\alpha^2},\\
& \propto& \frac{1}{k_x^2+k_y^2+\zeta^\prime\alpha^2+2\alpha p^{-1}_1(p_0k_x-q_0k_y)},\label{SEqn:QrD}
\eqa
where $\zeta^\prime=(p_0^2+q_0^2+\beta r_0^2)/p_1^2>0$. Since we are considering the scaling rule of the statistical average of $Q_r$ for a class of perturbations corresponding to the same parameter $\alpha$, the first moment of the independent $p$'s and $q$'s is zero, and thus the last term in the denominator of Eq. \ref{SEqn:QrD} vanishes. Thus, 
\bqa
Q_r &\propto&  \frac{1}{k_x^2+k_y^2+\zeta^\prime\alpha^2}\\
&\propto&  \left(\frac{n^2+m^2}{N^2}+\zeta\alpha^{2}\right)^{-1}.\label{SEqn:23}
\eqa
At the $\Gamma$ point, $Q_r \propto  \alpha^{-2}$, which is consistent with the result of Ref. \cite{koshelev2018asymmetric}. 

\subsection{Scattering loss}

\begin{figure}[htb]
	\centering
	\includegraphics[width=0.3\linewidth]{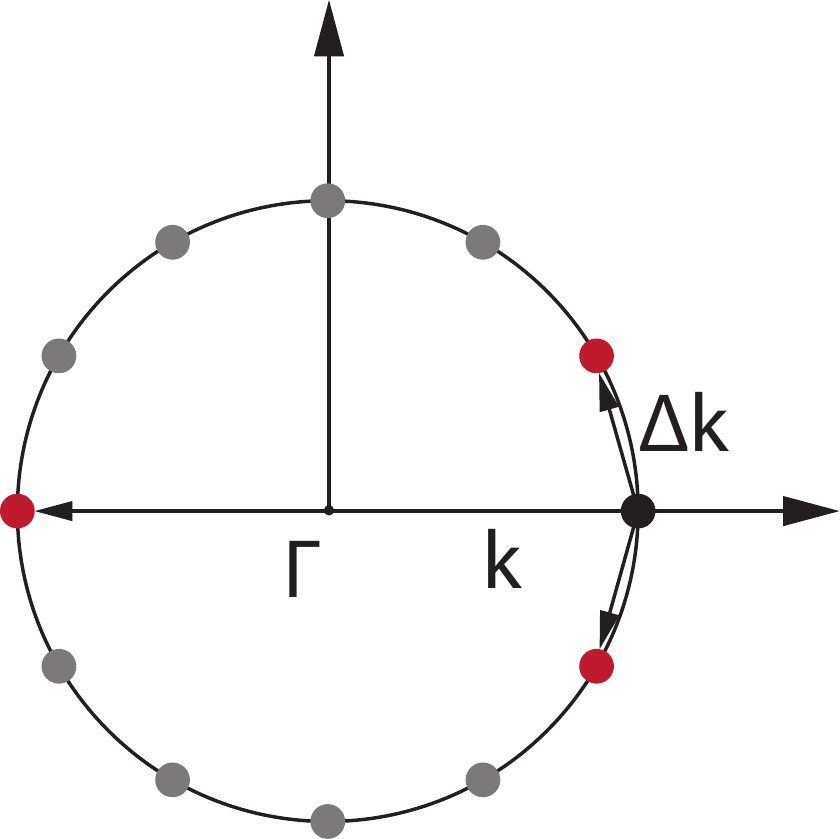}
	\caption{ Illustration of mode scattering on an isofrequency contour close to the $\Gamma$ point, dominated by scattering to the modes labeled by red dots. }
	\label{SFig:2}
\end{figure}

We provide a phenomenological analysis of the phonon scattering loss. The slight inhomogeneity of unit cells in actual PnCs causes weak scattering of phonons among isofrequency modes, which all have roughly the same radiative quality factor (Eq. \ref{SEqn:23}). As a consequence, the radiation loss of a band-edge mode is increased by a factor roughly equal with the number of isofrequency modes it can scatter to, according to Fermi's golden rule, which resembles Purcell's effect \cite{purcell1946spontaneous}. For relatively weak disorders, the dominant scattering process involves isofrequency modes in the vicinity of the mode under consideration because of large field-overlap and the mode with opposite momentum because of time-reversal symmetry induced coherent backscattering \cite{regan2016direct}, as shown in Fig. \ref{SFig:2}. As a result, the radiation enhancement factor is approximately an $O(1)$ constant $\lambda$ for the modes in the vicinity of the $\Gamma$ point, i.e., 
\be\label{SEqn:25}
\frac{1}{Q_r}+\frac{1}{Q_s}\approx \frac{\lambda}{Q_r}.
\ee

\subsection{Material losses}

Now we consider material-related losses and show they are all ignorable at the lowest temperature (60 mK) measured in this work, comparing to the acoustic radiation and scattering losses.\\

\emph{Phonon-phonon interaction}

At low temperature when $\omega_m\tau\gg1$, $\tau$ the thermal phonon relaxation time, the dominant phonon-phonon interaction is the Landau-Rumer effect, where acoustic phonons interact with individual thermal phonons. The effect is very weak though as experimentally showed, for example, in silicon optomechanical cavities \cite{chan2012laser, maccabe2019phononic}, leading to the characterizing $Q$ factor $Q_{a,\textrm{ph-ph}}>10^{13}$ below 1 K. We expect similar behavior of phonon-phonon interaction in AlN, with $Q_{a,\textrm{ph-ph}}$ at least a few orders larger than the measured $Q$ factor of the BIC standing-wave modes at low temperature ($\sim 10^4$).  \\

\emph{Grain-boundary relaxation}

The AlN used in this work is polycrystalline with grain size about 50 nm based on scanning electron microscopy. These grains are still pretty well aligned with x-ray diffraction analysis showing a rocking curve angle $< 1.5^\circ$. A natural speculation of another possible material loss is thus the grain-boundary relaxation. However, this effect is usually observed at elevated temperatures. Taking typical high purity metal as a reference, the grain-boundary relaxation activation energy $\Delta\epsilon\sim 0.1$ eV \cite{nowick1972anelastic} which is four orders larger than $k_{\mathrm{B}}T$ for $T=60$ mK, leading to ultralong relaxation time $\tau=\tau_0e^{\frac{\Delta\epsilon}{k_{\mathrm{B}}T}}$ ($\tau_0\approx10^{-12} - 10^{-10}$ s), irrelevant to the mechanical damping at such low temperature.\\

\emph{Two-level state}

Another common source of mechanical dissipation is via coupling between phonons and two-level states (TLS)--a generic defect state in a solid-state material which possesses two local arrangements of atoms with nearly degenerate energy and is associated with both electric and acoustic dipoles. TLS has been found as the dominant source for acoustic damping below 1 K in both single crystalline materials \cite{kleiman1987two, maccabe2019phononic} and vitreous glass \cite{phillips1987two}. For the polycrystalline AlN here, where TLS tends to aggregate on the grain boundary, we expect TLS-induced loss is also the dominant material-related acoustic loss at low temperature.

For high frequency, microwave phonon modes and TLS at cryogenic temperatures, and assuming dispersive phonon-TLS coupling, it is shown that the TLS-induced dissipation of a mechanical mode is given by \cite{phillips1987two,maccabe2019phononic}.
\begin{equation}\label{SEqn:TLS}
\gamma_{a, \mathrm{TLS}}=\sum_{\mathrm{TLS}}f\left[\mathbf{r}_{\mathrm{TLS}}\right]\frac{\omega_{\mathrm{TLS}}^{d}}{T} \operatorname{csch}\left[\hbar \omega_{\mathrm{TLS}} / k_{B} T\right],
\end{equation}
where $d$ is the dimension of the phonon bath, $\omega_{\mathrm{TLS}}$ is the frequency splitting of the TLS, and $f\left[\mathbf{r}_{\mathrm{TLS}}\right]$ is a function of the spatial location of TLS and parameters other than $T$ and $\omega_{\mathrm{TLS}}$, including TLS-phonon coupling rate, material density, and sound speed. For a uniform spectral and spatial density of states for the TLS, performing the integration in Eq. \ref{SEqn:TLS} yields
\begin{equation}\label{SEqn:TLS2}
\gamma_{a, \mathrm{TLS}}\propto n_{0,\mathrm{TLS}}V_mT^d
\end{equation}
where $n_{0,\mathrm{TLS}}$ is the number of TLS per frequency per volume and $V_m$ is the acoustic mode volume. 

First, we verify that the measured material-related quality factor of the BICs at 1 K is consistent with the TLS-induced loss in the AlN PnC by comparing with the result of Ref. \cite{maccabe2019phononic}. Ref. \cite{maccabe2019phononic} measured the $Q$ factor of a 5 GHz breathing mode of a silicon optomechanical crystal nanobeam to be $Q_{\textrm{Si}}\approx 2\times10^7$ at 1 K, and found it is limited by TLS-induced loss with $n_{0,\mathrm{TLS}}$=692.8 states/$\mu$m$^3$/GHz (taken from the value for the vitreous glass \cite{phillips1987two}) and etching-damaged, TLS-populated volume $V_{\textrm{Si}, m}=0.039$ $\mu$m$^3$. If we assume the TLS density for the polycrystalline AlN to be one order smaller than the vitreous glass (note single crystalline silicon is estimated to have TLS density two orders smaller than the vitreous glass \cite{kleiman1987two}), and consider the $B_1$ BIC with unit-cell mode volume $V_{\textrm{BIC},m}$=0.156 $\mu$m$^3$, then in a PnC with $200\times200$ unit cells, the TLS-loss limited quality factor is roughly $Q_a\approx\frac{V_{\textrm{Si}, m}}{200^2\times V_{\textrm{BIC},m}}Q_{\textrm{Si}}\approx10^3$, according to Eq. \ref{SEqn:TLS2}. This is actually close to the measured $Q_a$ for the BIC standing-wave modes at 1 K, which is $1/(1/Q_{1 \textrm{K}}-1/Q_{60 \textrm{mK}})\approx 10^4$, assuming $Q_{60 \textrm{mK}}$ is merely due to radiation and scattering loss which is to be verified right below. Now, taking the measured TLS-loss limited $Q_{a, \mathrm{TLS}}\approx 10^4$ at 1 K, using the scaling rule of Ref. \ref{SEqn:TLS2}, and given $d=3$ for our slab-on-substrate structure, we have
\be
Q_{a, \mathrm{TLS}}\approx 10^4\times\left(\frac{1\ \mathrm{K}}{60\ \mathrm{mK}}\right)^3\approx 4.6\times 10^7
\ee
at 60 mK, which is three orders larger than the measured $Q$ factor.

\subsection{Quality factor at base temperature}

Based on the analysis above, the quality factor of the mechanical BIC modes at 60 mK is dominated by the radiation and scattering losses, i.e.,
\be
\frac{1}{Q}\approx\frac{1}{Q_r}+\frac{1}{Q_s}+\frac{1}{Q_e}.
\ee
We also numerically confirmed that $Q_e>10^5$ for the fundamental mode in $y$-periodic PnCs with $N=50$ along the $x$-axis, and $Q_e$ increases with $N$, so it can be ignored comparing to the $Q$ factor measured at 1 K and 60 mK. Using these, we are able to determine the mode order of the three resonances of the PnC with 200$\times$200 unit cells observed at the base temperature. First, we assume the mode order along the $y$-direction to be $m=1$. This is because the $m=2$ modes are odd with respect to the $x-$axis and cannot couple with the incident wave, and the coupling of higher order even modes with the incident wave roughly scales as $1/m$ due to field oscillation along the $y$-axis, which leads to power transmission that scales as $1/m^2$, leaving these resonances too weak to be detected. Thus, we assume the mode order of the three resonances to be $(n, 1), (n+1, 1), (n+2, 1)$. According to Eqs. \ref{SEqn:23} and \ref{SEqn:25}, the ratio between the quality factor of these resonances is
\be
Q_n:Q_{n+1}:Q_{n+2}=\frac{1}{\frac{n^2+1}{N^2}+\zeta\alpha^{2}}:\frac{1}{\frac{(n+1)^2+1}{N^2}+\zeta\alpha^{2}}:\frac{1}{\frac{(n+2)^2+1}{N^2}+\zeta\alpha^{2}}.
\ee
Thus,
\be
\frac{Q_n}{Q_{n+1}}< \frac{(n+1)^2+1}{n^2+1}\leq 2,\quad n\geq 2
\ee
and
\be
\frac{Q_n}{Q_{n+2}}< \frac{(n+2)^2+1}{n^2+1}\leq  3.4, \quad n\geq 2
\ee
The measured values, however, are $\frac{Q_n}{Q_{n+1}}=2.5$ and $\frac{Q_n}{Q_{n+2}}=4.0$, which leads to the only option $n=1$. These lower order modes indeed have much weaker external coupling comparing to the measured radiation loss, consistent with the deduction above where $1/Q_e$ has been ignored.

\subsection{Simulation of radiation loss}

To simulate the radiation loss of BIC standing-wave resonances in structures with imperfection, we introduced variations of the dimension $b$'s and $c$'s in the unit cell. Each $b$ and $c$ in the unit cell is varied independently, constrained by percentage standard deviation $\alpha$. For each $\alpha$, we simulated 50 unit cell structures with randomly generated $b$'s and $c$'s, and by selecting the Bloch wavevector corresponding to the standing-wave mode, calculated the radiative quality factor of the latter.

The simulated radiative quality factor of $(1,1)$, $(2,1)$, and $(3,1)$ modes in the $200\times200$ PnC versus disorders, as shown in Fig. 4b, is then fitted globally using the formula
\be
Q_r=A\left(\frac{n^2+m^2}{N^2}+\zeta\alpha^{2}\right)^{-1}
\ee
with $A$ and $\zeta$ as the fitting parameters.

%%%%%%%%%%%%%%%%%%%%%%%%%%%%%%%%%%%%%%%%%%%%%%%%%%

\section{Absence of symmetry-induced BICs at $X$ and $M$ points}\label{App:C}

We prove that there are no symmetry-induced BICs at $X$ and $M$ points in $C_{4v}$ structures. 

The $X$ point has $C_{2v}$ symmetry, whose character table is provided in Table. \ref{tab:c2v}. Any Bloch mode at $X (\frac{\pi}{a}, 0)$ point can only possibly couple into two radiation channels, corresponding to plane waves with in-plane momentum $(\frac{\pi}{a}, 0)$ and $(-\frac{\pi}{a}, 0)$. The displacement field of these plane waves is given by
\begin{equation}\label{SEqn:X}
{\mathbf{Q}_{X}}(x,y) = (c_x^{(0,0)}\mathbf{\hat x} + c_y^{(0,0)}\mathbf{\hat y} + c_z^{(0,0)}\mathbf{\hat z}){e^{ + i\frac{\pi }{a}x}} + (c_x^{(-1,0)}\mathbf{\hat x} + c_y^{(-1,0)}\mathbf{\hat y} + c_z^{(-1,0)}\mathbf{\hat z}){e^{ - i\frac{\pi }{a}x}}.
\end{equation}
The three unit vectors $\mathbf{\hat x}$, $\mathbf{\hat y}$ and $\mathbf{\hat z}$ belong to the $B_1$, $B_2$, and $A_1$ representation, respectively. Meanwhile, $\cos (\frac{{\pi x}}{a})$ belongs to $A_1$ and $\sin (\frac{{\pi x}}{a})$ belongs to $B_1$. Thus, the representation of the two terms in Eq. \ref{SEqn:X} can be reduced as
\be
({B_1} + {B_2} + {A_1}) \otimes ({A_1} + {B_1}) = 2{A_1} + 2{B_1} + {A_2} + {B_2}.
\ee
As it contains all possible representations of the $C_{2v}$ group, any Bloch mode at the $X$ point thus will inevitably couple into these radiation channels and cannot be BICs.

The $M$ point has $C_{4v}$ symmetry, whose character table is provided in Table. \ref{tab:c4v}. Any Bloch mode at the $M$ point can only possibly couple into four radiation channels, corresponding to plane waves with in-plane momentum $(\pm\frac{\pi}{a}, \pm\frac{\pi}{a})$. The displacement field of these plane waves can be written as 
	\begin{equation}\label{SEqn:M}
	\begin{aligned}
	{{\bf{Q}}_M}(x,y) =&\ (c_x^{(0,0)}{\bf{\hat x}} + c_y^{(0,0)}{\bf{\hat y}} + c_z^{(0,0)}{\bf{\hat z}}){e^{ + i\frac{\pi }{a}x + i\frac{\pi }{a}y}} + (c_x^{( - 1,0)}{\bf{\hat x}} + c_y^{( - 1,0)}{\bf{\hat y}} + c_z^{( - 1,0)}{\bf{\hat z}}){e^{ - i\frac{\pi }{a}x + i\frac{\pi }{a}y}}\\
	&+ (c_x^{(0, - 1)}{\bf{\hat x}} + c_y^{(0, - 1)}{\bf{\hat y}} + c_z^{(0, - 1)}{\bf{\hat z}}){e^{ + i\frac{\pi }{a}x - i\frac{\pi }{a}y}} + (c_x^{( - 1, - 1)}{\bf{\hat x}} + c_y^{( - 1, - 1)}{\bf{\hat y}} + c_z^{( - 1, - 1)}{\bf{\hat z}}){e^{ - i\frac{\pi }{a}x - i\frac{\pi }{a}y}}.
	\end{aligned}
	\end{equation}
The three unit vectors $\mathbf{\hat x}$, $\mathbf{\hat y}$ and $\mathbf{\hat z}$ belong to $E$ and $A_1$ representations. Meanwhile, $\cos (\frac{{\pi x}}{a})\cos (\frac{{\pi y}}{a})$ belongs to the $A_1$ representation, $\sin (\frac{{\pi x}}{a})\sin (\frac{{\pi y}}{a})$ belongs to the $B_2$ representation, and $\cos (\frac{{\pi x}}{a})\sin (\frac{{\pi y}}{a})$ and $\sin (\frac{{\pi x}}{a})\cos (\frac{{\pi y}}{a})$ belong to the $E$ representation. Since $\exp ( \pm i\frac{\pi }{a}x \pm i\frac{\pi }{a}y)$ can be expressed as a linear combination of these four terms, the representation of the four terms in Eq. \ref{SEqn:M} can be reduced as
\be
(E + {A_1}) \otimes ({A_1} + {B_2} + E) = 2{A_1} + {A_2} + {B_1} + 2{B_2} + 3E.
\ee
Again, as they contain all possible representations of the $C_{4v}$ group, any Bloch mode at the $M$ point thus will inevitably couple into these radiation channels and cannot be BICs.

\begin{table}[htbp]

\caption{Character table for point group $C_{2v}$}
	\centering
	\begin{tabular}{c|c c c c}
		\hline
		\hline
		$C_{2v}$   & $E$     & $C_2$    & $\sigma_x$     & $\sigma_y$    \bigstrut\\
		\hline
		$A_1$    & 1     & 1     & 1     & 1      \bigstrut\\
		$A_2$    & 1     & 1     & -1    & -1    \bigstrut\\
		$B_1$    & 1     & -1    & 1     & -1     \bigstrut\\
		$B_2$    & 1     & -1    & -1    & 1   \bigstrut\\
		\hline
	\end{tabular}%
	\label{tab:c2v}%
\end{table}%

\begin{table}[htbp]
	\caption{Character table for point group $C_{4v}$}
	\centering
	\begin{tabular}{c |c c c c c} 
		\hline\hline
		$C_{4v}$ & $E$ & $2C_4$ & $C_2$ & $2\sigma_v$ & $2\sigma_d$ \\ %[0.5ex] % inserts table
		%heading
		\hline % inserts single horizontal line
		$A_1$ & 1 & 1 & 1 & 1 & 1 \\ % inserting body of the table
		$A_2$ & 1 & 1 & 1 & -1 & -1\\
		$B_1$ & 1 & -1 & 1 & 1 & -1\\
		$B_2$ & 1 & -1 & 1 & -1 & 1\\
		$E$ & 2 & 0 & -2 & 0 & 0 \\ %[1ex] % [1ex] adds vertical space
		\hline
	\end{tabular}
	\label{tab:c4v}
\end{table}
 
\end{document}